\documentclass[a4paper]{article}
\usepackage{latexsym}

\usepackage[english]{babel}
\usepackage[utf8]{inputenc}
\usepackage{textcomp,color}
\usepackage{graphics,graphicx}
\usepackage{hyperref}
\usepackage{overpic}
\usepackage{graphicx}
\usepackage{amsmath}
\usepackage{amsfonts}
\usepackage{amssymb}

\title{Distortion of extra dimensions \\ in the inflationary Multiverse}

	\author{
	Sergey G. Rubin$^{1,2,a}$  \and
J\'ulio C. Fabris$^{1,3}$}

\date{%
$^1$  National Research Nuclear University MEPhI \\(Moscow Engineering Physics Institute), \\ 115409, Kashirskoe shosse 31, Moscow, Russia \\
$^2$N.I. Lobachevsky Institute of Mathematics and Mechanics,\\
	Kazan  Federal  University, \\
	Kremlevskaya  street  18,  420008  Kazan,  Russia\\
$^a$	sergeirubin@list.ru  \\
$^3$ 	N\'ucleo Cosmo-ufes \& Departamento de F\'{\i}sica,\\ Universidade Federal do Esp\'{\i}rito Santo, Vit\'oria,\\ ES, CEP 29075-910, Brazil}

\begin{document}
\maketitle

\begin{abstract}
We discuss the effect of the quantum fluctuations at high energies on the final shape of compact extra dimensions. The quantum fluctuations produce a wide range of the initial extra metrics in causally disconnected regions (pocket universes) of the Multiverse during the inflationary stage. This set of initial extra metrics evolves to a set of inhomogeneous metrics at the present time. The low energy physics appears to be different in different pocket universes.
The numerical estimate of the probability of finding a specific metric is based on the model of the compact 2-dimensional extra space.
\end{abstract}

\section{Introduction}

The inflationary paradigm has been developed decades ago, starting with the papers \cite{Starobinsky:1980te,Guth:1980zm}. The existence of an inflationary period in the
history of our Universe seems inevitable since it explains the key problems of the Big Bang theory: the horizon problem,
the flatness problem, the galaxies formation etc. \cite{Lindebook,KhlopovRubin}. The first inflationary
mechanisms were based on self-consistent equations
of the scalar and gravitational fields.  The key point is that the self gravitating scalar field leads to an exponential growth of initially small spatial regions at high energies. The horizon size appears to be much smaller then the expanding region. Therefore, the latter is separated into quickly increasing number of the causally disconnected regions. The estimation made by A. Linde \cite{Lindebook} gives $\mathfrak{N}\sim 10^{10^{12}}$ such regions -
pocket universes (PU) \cite{Guth:1997wk} after the inflation is finished.

Immediate question arises - are the physical parameters equal in the different PU? Positive answer open the way for solution of the fine tuning problem - one of the most serious enigma of modern physics to our mind. To make physical laws different in the different PU, we need a substantial variety of ground states. Nevertheless, it seems that the PU have similar properties for simple potentials. Even the idea of landscape does not improve the situation without knowledge of the ground states distribution. 

In this paper we discuss the way of endowing each PU by a specific ground state. To this end, we employ the idea of the compact extra dimensions. The extra dimensions are widely used tool for study deep problems of the modern physics. 
In particular, the physical parameters $h_i$ of the Standard Model depend on the extra space metric at low energies after the extra coordinates have been integrated out.


It is usually assumed that an extra space metric should satisfy generalized Einstein differential equations. Hence, there is a continuum set of solutions that differs by arbitrary constants. The choice of their specific values is usually made by comparison of the observations with the model predictions.  As was shown in  \cite{Rubin:2015pqa,Bronnikov:2020tdo}, the simplest metric is not necessarily favorite. 

Let us mention several examples for illustration. In the paper \cite{Goldberger:1999uk}, arbitrary constants are fixed by energetic argument which is not very strong one if the gravity is included, see discussion in \cite{2002PhRvD..66b4036C}. Very often, like in the paper \cite{Guo_2018}, the constants used to restore the 4-dim Planck mass and the cosmological constants.
The classical solutions explicitly containing two arbitrary constants in 5-dim brane world are considered in \cite{Bazeia:2013uva,Hashemi_2018}
A flat, two-dimensional toroidal compactification is discussed in \cite{Dienes:2001wu}.
where the relative angle between the two internal directions can be continuously varied to alternate the extra space volume. 

We conclude that there is an enormous number $\mathfrak{N}$ of separate universes - PU - on the one hand, and a continuum set of different extra metrics, on the second hand. The idea is to relate different extra metric and different PU. 

The discussion above does not specify extra space metric and topology. Here, we focus on particular set of the compact extra spaces characterized by inhomogeneous extra metrics. The latter has been revealed in \cite{Rubin:2015pqa} and their stability was discussed in \cite{Bronnikov:2020tdo}. Their application leads to variety of effects, see  \cite{1983PhRvL..51..931A,2002PhRvD..65b4032A,Gani:2014lka,Nikulin2019}. Such kind of metrics is used in the other research like the Universal Extra Dimensions approach \cite{ArkaniHamed:1998rs}, the branes formation \cite{Rubin:2015pqa} the justification of the 3 generation of leptons \cite{Gogberashvili_2007} and the baryon asymmetry of the Universe \cite{Rubin:2018ybs}. 


In this letter, we pay attention to the way of such metrics formation. We show here that the quantum fluctuations are responsible for the final shape of metric. It is known that the scalar field tends to a potential minimum if the gravity is not involved into consideration. It seems that nothing changed if we add maximally symmetric extra dimensions. Any scalar field excitation acting in the extra dimensions (like KK-modes) also is attenuated sooner or later.
The picture is changed drastically if the gravity is taken into account.
Indeed, the energy density of fluctuations in the FRW universe evolves into dense local objects forming the large scale structure of the Universe. In this letter, we show that the gravity influences the scalar field fluctuations within the compact extra dimensions in the same manner. The quantum fluctuations produced at the inflationary stage deform the extra space metric that tends to stationary state after the end of inflation.



In Section II we describe the basis of the following discussion - the $D$-dimensional scalar-tensor model with arbitrary function $f(R)$. 4-dim space is endowed by the de Sitter metric. Section III is devoted to stationary states in extra dimensions mainly based on paper \cite{Bronnikov:2020tdo}. The classical states are defined up to the quantum fluctuation which could be large at high energies.  In Section IV we elaborate the way to work with the scalar field quantum fluctuations in $4+n$ dimensional space. In Section V we estimate the formation probability of specific pocket universes.



\section{Outline}
Consider the manifold with the topology $M_1\times M_3\times M_n$. The space $M_1\times M_3$ - is described by the de-Sitter metric, $M_n$ - represents the ideal $n$-dimensional sphere, its deformation is described below. Our model is based on the action
\begin{eqnarray}                           \label{S}
	    S &=&  \int d^D x \sqrt{|g_D|} \bigg[ \frac{m_D^{D-2}}{2} f(R)
	     +  \frac 12 g^{AB} \phi_{,A}\phi_{,B} - V(\phi) \bigg],  \\
       && A, B = 0, \ldots, 5. \nonumber
\end{eqnarray}  	
  where $g_D = \det (g_{AB})$, $D=4+n$, $n$ is a number of extra dimensions.  $f(R)$ and $V(\phi)$ are some functions of the D-dim scalar curvature $R$ and the scalar field $\phi$, respectively. Requirements for the potential are quite easy and we choose the simplest form $V(\phi)=m^2\phi^2/2$.
  Variation of action $S$ with respect to $\phi$ and $g^{AB}$ leads to the field equations
\begin{eqnarray}\label{Eqphi}
    	    \square \phi +  V_\phi =0, \quad {\rm where}\ 
	    \square\phi = \nabla_A \nabla^A \phi, \quad  V_\phi = dV/d\phi,	\label{eq-phi}
\end{eqnarray}

\begin{eqnarray}\label{EE}
	   && -\frac 12 \delta_A^B f (R) + \big[ R^B_A + \nabla_A \nabla^B - \delta^B_A \square \big] f_R
	        = - \frac{1}{m_D^2} T^B_A, \\
	   && \hskip3mm f_R = df/dR, \nonumber
\end{eqnarray}	  		  		  		  		  		  		  		
Let it be pure de-Sitter 4-dim space.
Then the stress-energy tensor of the scalar field $\phi(t,u)$ reads
\begin{eqnarray}\label{TAB}
		T^B_A[\phi] = \phi_{,A}\phi^{,B} - \frac 12 \delta^B_A \phi_{,C}\phi^{,C} + \delta^B_A V.
\end{eqnarray}
The interval
\begin{equation}\label{metric}
	ds^2 = dt^2 - e^{2 H t} \delta_{i j} dx^i dx^j - du^2-r(u)^2 d\varphi^2
\end{equation}
for the 2-dim compact extra space $(n=2)$ with the sphere topology and $H=const$.
Here $i,j=1,2,3$ and the extra space coordinates $y=(u,\varphi)$.

 As was shown in \cite{Rubin:2015pqa, Bronnikov:2020tdo}, system \eqref{EE} is reduced to simpler system 
\begin{eqnarray}           \label{phi''}
	 &&\phi'' + \phi'\frac{r'}{r}  = V'_\phi,	\\
&&                 \label{EE00}
 	  -\frac12 f(R) +3 H^2 f_R + f_R'' +\frac{r'}{r} f_R' = m_D^{-2}\left( -\frac{\phi'^2}{2} - V \right),
\\&&
                 \label{EE44}
          -\frac12 f(R) - \frac{r''}{r} f_R +\frac{r'}{r} f_R' = m_D^{-2} \left( \frac{\phi'^2}{2} - V \right),
\\&&
		      \label{EE55}
	 -\frac12 f(R) - \frac{r''}{r} f_R + f_R'' = m_D^{-2}\left( -\frac{{\phi'}^2}{2} - V \right),    
\end{eqnarray}
for metric \eqref{metric} and the stationary scalar field depending on the internal coordinate $u$, $\phi = \phi(u)$. Here $f'_R = df_R/du$, etc. The system described above is the basis for the following discussion.

The Ricci scalar is expressed in terms of the metric functions as
\begin{equation}\label{R}
    	R = 12 H^2 - \frac{2 r''}{r}
\end{equation}
where $'\equiv d/du$.

Throughout this letter, we use the conventions for the curvature tensor $R_{\mu\nu \alpha}^{\beta}=\partial_{\alpha}\Gamma_{\mu \nu}^{\beta}-\partial_{\nu}\Gamma_{\mu \alpha}^{\beta}+\Gamma_{\sigma \alpha}^{\beta}\Gamma_{\nu\mu}^{\sigma}-\Gamma_{\sigma \nu}^{\beta}\Gamma_{\mu \alpha}^{\sigma}$ and the Ricci tensor is defined as $R_{\mu \nu}=R^{\alpha}_{\mu \alpha \nu } \, $.

\section{Static inhomogeneous extra space}

\subsection{f(R) gravity without a matter}\label{fR}

In this section, we follow the analysis of system \eqref{EE} made in \cite{Bronnikov:2020tdo}. More definitely, the difference \eqref{EE00}-\eqref{EE44} leads to equation
\begin{eqnarray}
          \label{EE05}
		3H^2 + \frac{r''}{r} =0
\end{eqnarray}
for the de Sitter space with the Hubble parameter $H$.
Here we have kept in mind that $\phi = 0$, $f_R\neq 0,\, R=const$ and hence $R'=0$. The regular center is characterized by conditions $r(0)=0, r'(0)=1$. Together with equation \eqref{EE05}, these conditions give the solution 
$$r(u)=r_0 \sin(u/r_0).$$
with the constant radius $r_0$ of extra dimensions depending on $H$,
\begin{equation}\label{lH}
    r_0=\frac{1}{\sqrt{3}H} 
\end{equation}
Expressions \eqref{R} and \eqref{EE05} give
the 6-dim Ricci scalar
\begin{equation}\label{RH}
    		 R = 18 H^2 
\end{equation}
The 2-dim metric has the standard form
\begin{equation}\label{l0}
    ds_2=r_0^2[dv^2+\sin^2(v)d\varphi^2]
\end{equation}
in terms of new coordinate $v=u/l_0$. This final metric does not contradict our assumption $R=const$ and gives the relation \eqref{lH} between the Hubble parameter and the extra dimensional radius $r_0$. At the present time $H\simeq 0$ and hence $r_0\to \infty$ { which strongly contradicts observations. As has been shown in \cite{Rubin:2015pqa,Bronnikov:2020tdo}, the inhomogeneous compact extra dimensions are free from this defect. Here we study one of the way of such metrics formation}. 

At high energies, the maximally symmetrical extra space does exist. Its Ricci scalar is proportional to the Hubble parameter square. 
Below we consider influence of the scalar field distribution to these basic metric.

\subsection{Continuous set of the extra dimensional metrics. The scalar field in the D-dim space.}

Let us decompose the scalar field in action \eqref{S} in the standard manner 
\begin{equation}\label{sdec}
	\phi(x,y)=\sum_a \Phi_a (x)Y_a (y)
\end{equation}
where the orthonormal functions $Y_a (y)$ satisfy the equations 
\begin{equation}\label{eigen}
	\square_{n}Y_a=\lambda_a Y_a, \quad 2\pi\int_0^{u_{max}}|Y_a|^2\sqrt{|g_n|}du =1.
\end{equation}
Here $n$ is the dimensionality of the extra space and we assume the homogeneity in $x$ directions. The extra dimensions are described by $u,\phi$ coordinates. The upper limit of the integral can be found in the capture to Fig.1.
Following the Kaluza-Klein approach, one can substitute decomposition \eqref{sdec} into \eqref{S} to obtain an action as a sum of scalar modes. The destiny of these scalar modes acting in the 4-dim de Sitter universe is well known - they are attenuated according to the equation
\begin{equation}\label{ddphi}
	\ddot{\Phi}_a+3H\dot{\Phi}_a+\mu^2_a\Phi_a=0;\quad \mu_a^2 =m^2 + \lambda_a 
\end{equation}
for the quadratic potential $V(\phi)={m^2} \phi^2/2$.
The final state is common stationary state $\Phi_a = 0$ for all modes. Therefore, we are back to the situation described above in subsection \ref{fR}. The scalar field equals zero and the only asymtote of the 2-dim metric is the maximally symmetric one with positive curvature.

The picture changes drastically when we keep in mind a deformation of the extra space metric under the scalar field influence. In this case, we have to solve the system of equations \eqref{EE}, rather than single equation \eqref{ddphi}. This is a difficult task, and we will restrict ourselves to discuss the asymptotes for $t\to\infty$.

\begin{figure}
\centering
\includegraphics[width=9cm]{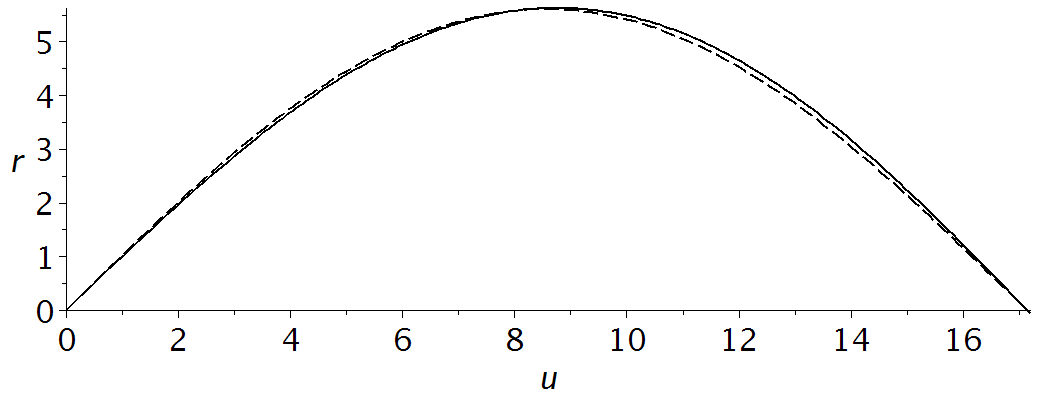} \quad \includegraphics[width=9cm]{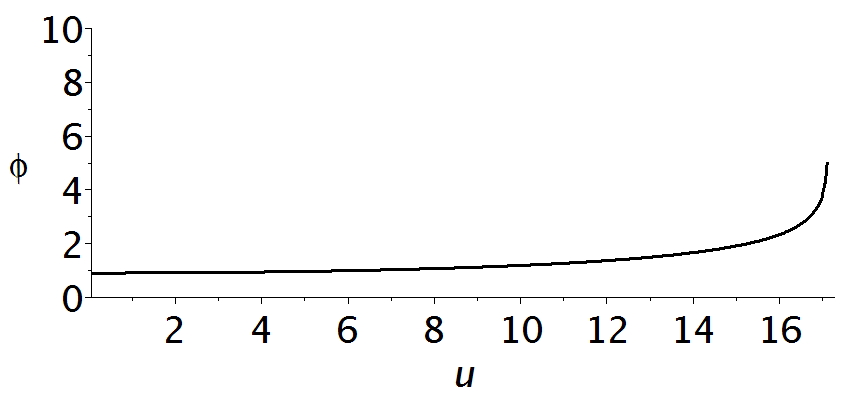}
\caption{\small
	The extra space metric function $r(u)$ (solid line, upper figure) and the scalar field $\phi(u)$ (lower figure)
	for $f(R)=aR^2 +bR +c$ and $V(\phi)=(m^2/2) \phi^2$ (units $m_D=1$). Dashed line relates to the maximally symmetric 2-dim sphere. The parameter values are
	$ m=0.1, \ b=1, \ a=-10.9, \ c=-0.0021$. Additional conditions are: $\phi_0=0.9, \ H=0.1$
	and $R(0) \simeq 0.182$. The variation range of the parameter $u$ is $(0\div u_{max}),\, u_{max}=17.1.$}
\label{Heq01}\end{figure}

There is infinite number of stationary solutions to system  \eqref{phi''}, \eqref{EE00}, \eqref{EE44}, \eqref{EE55} depending on the additional conditions \cite{Rubin:2015pqa,Bronnikov:2020tdo}. 
It was revealed there that the continuum set of nontrivial stationary solutions does exist. Each solution is characterized by a specific additional condition. For illustration, we represent one of the stationary solution of the system, see Fig.\ref{Heq01}.

As was shown in  \cite{Rubin:2015pqa, Bronnikov:2020tdo}, each stationary metric relates to a specific value of the cosmological constant.  The maximally symmetric 2-dim extra space is one of the solutions. The latter is unacceptable because of the Hubble parameter smallness at the present time. Indeed, the observable value $H_0\sim 10^{-42}GeV$ that gives too big size \eqref{lH} of extra dimensions comparable to the Universe size.

Fortunately, there is infinite set of the stationary states. As one can see from the Figure \ref{Heq01}, upper panel, that the size of distorted extra space is quite small, $\sim 5$ in $m_D$ units. As was revealed in \cite{Rubin:2015pqa,Gani:2014lka,Bronnikov:2020tdo}, there exists a solution to the system \eqref{phi''} - \eqref{EE55} for which the extra space size remains small even if $H=0$. So, the observable value of the cosmological constant and small size of the extra space may not contradict each other. 

Each area under the horizon - PU - can be equipped with the specific extra space metrics. It means that different universes are characterized by different cosmological constants. 

Before moving further, an important remark is necessary.
The quantum fluctuations $\delta\phi$ play important role at the inflationary stage when the Hubble parameter is large. It is assumed that the scalar field consists of classical and quantum parts
\begin{equation}\label{cl+q}
\phi(t,u)=\phi_{cl}(u)+\delta\phi (t,u).
\end{equation}
The classical part $\phi_{cl}$ is good approximation if the quantum fluctuations are small,
\begin{equation}\label{gg0}
\sqrt{<\delta\phi^2>}\ll \phi_{cl}(u).
\end{equation}
This condition is considered true when the stationary solutions of the system \eqref{phi''}, \eqref{EE00}, \eqref{EE44}, \eqref{EE55} are discussed.

\section{Scalar field fluctuations in D-dim space}

The previous section is devoted to the discussion on the inhomogeneous extra dimensional metrics. They form continuous set of stable metrics. Now we will consider the way they are formed.

Analogy to the 4-dim collapse of dust to the black hole is good support for our intuition. Indeed, the specific black hole mass is the result of initial dust distribution which collapses to this black hole. In its turn, the collapsing energy density distribution is the result of the quantum fluctuations at the inflationary stage.
Evidently, the resulting black hole masses form a continuous set.
The quantum fluctuations of scalar field are also a reason of the continuum set of inhomogeneous extra metrics.

Important question is the probability of appropriate quantum fluctuations. The subsequent mechanism has been substantially studied in the inflationary scenario \cite{KhlopovRubin} and we use it below. 

\subsection{4-dim fluctuations}

This subsection is needed to remind results for the standard 4-dim inflation which will be used later.

Let us start with the scalar field of mass $m$ and the standard action \eqref{S} in 4 dimensions.
It is known that the quantum fluctuations of scalar field quickly acquires classical behavior described by equation \eqref{Eqphi}. The latter depends on metric but not on the same function $f(R)$. Hence, we may use known results for the scalar field fluctuations acting in the de Sitter metric.
We are interested in the super horizon scales for which the fluctuations do not depend on the space coordinates.
The probability to find the field value $\phi_2$ at the instant $t_2=t_1+t$ in a space region of the horizon size $H^{-1}$ is, see \cite{KhlopovRubin}
\begin{equation}\label{Prob}
	dP(\phi_2,t_1+t;\phi_1,t_1|m)=d\phi_2 \cdot\sqrt{q/\pi}\exp\left[-q(\phi_2 -\phi_1e^{-\mu t})^2\right]
\end{equation}
where
\begin{eqnarray}\label{key}
	&&q=\frac{\mu}{\sigma^2}\left(1-e^{-2\mu t}\right)^{-1},\\
	&&\mu=\frac{m^2}{3H}=const, \\
	&&\sigma = \frac{H^{3/2}}{2\pi}=const. 
\end{eqnarray}

The average field value can be derived from \eqref{Prob}
\begin{eqnarray}\label{corr}
&&	<\phi(t)^2>=\int_{-\infty}^{\infty}\phi^2dP(\phi,t_1+t;\phi_1,t_1|m) \\
&&=\phi_1^2 e^{-2\mu t}+\frac{3H^4}{8\pi^2 m^2}\left(1-e^{-2\mu t}\right). \nonumber
\end{eqnarray}
This expression coincides with that known in the literature  \cite{Rey:1986zk}. The first term relates to the classical motion in the 4-dim space while the second one is responsible for the quantum fluctuations.

\subsection{D-dim fluctuations}
In this subsection, we extend the results discussed above to the $D$-dim model with metric \eqref{metric} and action \eqref{S}. In the spirit of the Kaluza-Klein approach, the scalar field can be decomposed in the series \eqref{sdec} of the orthogonal normalized functions $Y_a$.
The discrete set of eigen-values $\lambda_a\, a=0,\pm 1,\pm 2...$ depends on the extra space metric. Here remark is necessary. The analytical manipulations are much more clear if the extra metric is maximally symmetric. Fortunately, the metric satisfying the system of equations is almost maximally symmetric indeed as can be seen from Fig.\ref{Heq01}, upper panel, at least for some additional conditions. 

The scalar part of the action acquires the form 
\begin{equation}\label{Ss}
	S=\sum_a \frac 1{2}\int d^{4}x\sqrt{g_{4}}[(\partial\Phi_a)^2 -  \mu_a^2\Phi_a^2],\quad \mu_a^2=m^2+\lambda_a.
\end{equation}
where decomposition \eqref{sdec} has been used.

Now, formulas \eqref{Prob}-\eqref{corr} can be applied to each mode $\Phi_a$. In particular,
\begin{eqnarray}\label{key2}
&&<\phi(t)^2>=\sum_{a=-\infty}^{\infty}<\Phi_a(t)^2> = \sum_{a=-\infty}^{\infty} \frac{3H^4}{8\pi^2\mu_a^2},\quad t\to \infty
\end{eqnarray}
 according to \eqref{corr}. This value characterises the amplitude of fluctuations.

The fluctuation probability for first $A$ modes $\Phi_a,\, (a=0,\pm 1, \pm 2,..,\pm A)$ can be found as (see \eqref{Prob} for $t\to \infty$)
\begin{eqnarray}\label{Probc}
&& dP=\prod_{a=-A}^{A} dP_a= \prod_{a=-A}^{A} d\Phi_a \cdot\sqrt{q_a/\pi}\exp\left[-q_a\Phi_a^2 \right],\quad 
\end{eqnarray}



\section{Formation of extra space metrics by quantum fluctuations}

Let we have compact $n$-dim maximally symmetric extra space \eqref{l0} of the radius $r_0$. The classical form of metric is defined by system  \eqref{phi''} - \eqref{EE55}  approximately, up to the quantum fluctuations, see \eqref{gg0}. 
Some fluctuations could serve as the initial conditions of system  \eqref{phi''} - \eqref{EE55} if their amplitudes are large with respect to the average value $<\phi>$.
There is a nonzero probability \eqref{Probc} of such large amplitude scalar field fluctuation $\Phi_a$,
\begin{eqnarray}\label{gg}
\Phi_a \gg \sqrt{<\phi^2>},\quad  a=0,\pm 1, \pm 2,..,\pm A
\end{eqnarray}
Being created, it participates in the classical dynamics, influences both the metric and the field evolvution to a stationary state, the solution  to general system  \eqref{phi''} - \eqref{EE55}. Stationary $n$-dim extra space metrics form a set of the cardinality of the continuum, as discussed in the Introduction and studied in \cite{Rubin:2015pqa,Gani:2014lka,Bronnikov:2020tdo}. The realisation of a particular stationary metric depends on the initial conditions - an accidental set of fluctuations $\{\Phi_a\}$. The latter forms an initial scalar field distribution
\begin{eqnarray}\label{decompin}
\phi_{in}(u) \equiv \sum_{a=-A}^{A}\Phi_a Y_a(u)
\end{eqnarray}
Here $A$ is finite number. The equality $\phi_{-A}=\phi_{A}^*$ is assumed because the scalar field is real. The probability of such fluctuation contains 
the interval $d\Phi_a$ which also can not be smaller than the average quantum fluctuation $\sqrt{<\phi^2>}$. Otherwise, the fluctuations quickly push the field out from the desired interval. For estimations, we suppose that
\begin{equation}\label{dPhia}
    d\Phi_a\simeq \Delta \Phi_a\sim \sqrt{<\phi(t)^2>} \simeq \sqrt{\sum_{a=-A}^{A} \frac{3H^4}{8\pi^2\mu_a^2}}
\end{equation}
The probability of first $A$ modes fluctuation can be found as (see \eqref{Probc})
\begin{eqnarray}\label{Proba}
&&\Delta P \simeq 
 \prod_{a=-A}^{A}\sqrt{<\phi^2>}\cdot \sqrt{q_a/\pi}\cdot\exp\left[-q_a\Phi_a^2 \right]; \\
&& q_a=4\pi^2\frac{\mu_a}{H^3}.\nonumber
\end{eqnarray}

Now we can estimate the probability to find the scalar field distribution $\phi_{in}(u)$ in a specific point of our 3-dim space. It will help us to evaluate a number of pocket universes containing the particular deformed extra space. 
It is known that the masses of the Kaluza-Klein excitations are proportional to the inverse size of extra dimensions, $\lambda_a \sim 1/r$. Hence, the approximation $\mu_a\sim 1/r$ looks reasonable and we come to the following approximate equality, see \eqref{dPhia}:
$$\sqrt{<\phi^2>}\sim \sqrt{\frac{3A}{4\pi^2}}H^2r$$
$$q_a \sim\frac{4\pi^2}{H^3 r}$$
We have to consider modes with large amplitudes, see \eqref{gg}, so that
\begin{equation}\label{Phia}
    \Phi_a =\xi \cdot \sqrt{<\phi^2>}\sim\xi\cdot \sqrt{\frac{3A}{4\pi^2}}\cdot{H^2}{r},\quad \xi \gg 1%
\end{equation}
{The parameter $\xi$ characterises the ratio of first modes as compared to the averaged fluctuation and should be big in our case.}
Suppose that the modes number $A = 5$ which looks enough for suitable approximation. The choice $\xi = 10, H=0.1, r=10$ is discus below.
Substitution these values into \eqref{Proba} gives an estimation for a fraction of universes 
\begin{equation}\label{Probf}
\frac{\Delta N}{N}=\Delta P\sim e^{-6A^2\xi^2 Hr}\sim e^{-12500}\sim 10^{-5428}
\end{equation} 
containing such fluctuations. Here pre-exponent is neglected. This fraction is extremely small, but it does not mean that a number $\Delta N$ of such pocket universes is also small. Indeed, a total number $N$ of pocket universes - the causally disconnected domains under the present horizon of the Multiverse - can be estimated on the basis of the chaotic inflation \cite{Lindebook} as
\begin{equation}\label{Ntot}
    \mathfrak{N}\sim 10^{10^{12}}
\end{equation}
for the quadratic potential $V(\phi)={m^2} \phi^2/2$. This number is much greater than the number $10^{500}$ of the pocket universes usually mentioned in connection to the string theory.
Therefore, the number of such pocket universes
\begin{equation}\label{DN}
        \Delta N\sim \Delta P\cdot \mathfrak{N}\sim  10^{10^{12}}
\end{equation}
is really huge if one takes into account the total number of pocket universes \eqref{Ntot}.

It is also of interest to find such
a truncated set $\tilde{\Phi}_a$, the evolution of which is finished by a particular stationary solution $\phi_{cl}(u)$ of equations \eqref{phi''}-\eqref{EE55}. The quantum fluctuations prevent the exact solution of this task. Nevertheless, approximate way is evident. To this end, one has to approximate the required distribution  $\phi_{cl}(u)$ by the truncated set $\tilde{\Phi}_a, \, a=1,2,..,A$.
\begin{eqnarray}\label{decomp}
\phi_{cl}(u)\simeq \tilde{\phi}_{in}(u)=\sum_{a=-A}^{A}\tilde{\Phi}_a Y_a(u)
\end{eqnarray}
to obtain required modes
\begin{eqnarray}\label{phia}
\tilde{\Phi}_a =2\pi\int_0^{u_{max}}\tilde{\phi}_{in}(u)Y_a(u)\sqrt{|g_n|}du,\quad |a|\leq A
\end{eqnarray}
and find the probability according to \eqref{Proba}. Let us roughly estimate it for the stationary state represented in Fig. \ref{Heq01} where necessary numerical values can also be found. In fact, it has been done just above, see formula \eqref{Probf}-\eqref{DN}. The only what remains is to explain the estimation $\xi =10$ for the multiplier $\xi$ defined in \eqref{dPhia}. This estimation looks reasonable because $\sqrt{<\phi^2>}\sim 0.1$ (see last equality in  \eqref{dPhia}) and $\Phi_a \sim 1$ for the first several modes of the scalar field distribution in Fig.\ref{Heq01}.
As the result, we have more than enough amount of appropriate pocket universes with chosen truncated set of modes $\Phi_a$. This set eventually goes to a stationary state, similar to that shown in the Fig. \ref{Heq01}.

We would like to stress once again that the initial field distribution $\tilde{\phi}_{in}$ can evolve to a variety of slightly different stationary metrics $\phi_{cl}(u)$ under the influence of the quantum fluctuations. All of them have similar probability to be formed.

%





We can conclude that there are regions of the horizon size where the amplitudes of specific modes are quite large as compared to the quantum fluctuations. These modes are stretched in 3 dimensions.  
Simultaneously, such fluctuations influence the extra space metric and both of them evolve classically to stationary states.

\section{Conclusion}
The extra space paradigm is a powerful tool in the modern theoretical physics. They can influence low energy physics by determining the parameters $h_i$ of the Lagrangian.
The latter are fixed or limited by observations, which gives an indirect way of choosing a specific metric of extra space.  As has been shown in \cite{Bronnikov:2020tdo}, there exists a continuous set of the extra metrics. The measure of maximally symmetrical metrics appears to be extremely small.

In this letter, we discuss the mechanism of various extra space metrics formation. It is shown here that the quantum fluctuations of scalar field during the inflation are the reason of its nontrivial distribution over the compact extra dimensions. In its turn, this inhomogeneous distribution acts on the extra space metric leading to formation the brane-like structures studied in \cite{Bronnikov:2020tdo}. The way to estimate the probability of finding the specific extra metric $g_n$ in a space under the present horizon is developed.
Each volume under the present horizon (the pocket universe) is characterised by a specific value of the extra metric. 

The physical parameters $h_i$ depend on the extra space metric $g_n$ after the extra coordinates are integrated out. We come to conclusion that each pocket universe is endowed by a specific set of the physical parameters $h_i (g_n)$. A subset of such pocket universes could contain the observable parameters. The measure of such universes is estimated to be small but nonzero, which is enough to an intelligent life nucleated.


\section{Acknowledgments}
The work of SGR has been supported by the Ministry of Science and Higher Education of the Russian Federation, Project “Fundamental properties of elementary particles and cosmology” N 0723-2020-0041 and the Kazan Federal University Strategic Academic Leadership Program. JCF thanks CNPq (Brazil) and FAPES (Brazil) for partial financial support.


\end{document}